\documentclass[%
 reprint,
 amsmath,amssymb,
 aps,prl
]{revtex4-1}

\usepackage{graphicx}
\usepackage{dcolumn}
\usepackage{bm}
\usepackage{braket}
\usepackage{color,soul}
\usepackage{mathtools}
\usepackage{commath}

\begin{document}


\title{Characterization of coherent quantum frequency combs using electro-optic phase modulation}

\author{Poolad Imany}
\author{Ogaga D. Odele}

\author{Jose A. Jaramillo-Villegas}
\thanks{Also affiliated with Facultad de Ingenier\'{i}as, Universidad Tecnol\'{o}gica de Pereira, Pereira, Risaralda 660003, Colombia}

\author{Daniel E. Leaird}
\author{Andrew M. Weiner}
\email{amw@purdue.edu}
\affiliation{School of Electrical and Computer Engineering, Purdue University, West Lafayette, Indiana 47907, USA}
\affiliation{Purdue Quantum Center, Purdue University, West Lafayette, Indiana 47907, USA}

\date{\today}

\begin{abstract}
We demonstrate a two-photon interference experiment for phase coherent biphoton frequency combs (BFCs), created through spectral amplitude filtering of biphotons with a continuous broadband spectrum. By using an electro-optic phase modulator, we project the BFC lines into sidebands that overlap in frequency. The resulting high-visibility interference patterns provide an approach to verify frequency-bin entanglement even with slow single-photon detectors; we show interference patterns with visibilities that surpass the classical threshold for qubit and qutrit states. Additionally, for the first time, we show that with entangled qutrits, two photon interference occurs even with projections onto different final frequency states. Finally, we show the versatility of this scheme for weak-light measurements by performing a series of two-dimensional experiments at different signal-idler frequency offsets to measure the dispersion of a single-mode fiber.
\end{abstract}

\maketitle
\section{I. Introduction}
The desire to execute computationally complex algorithms in polynomial time and for complete security in communication networks has led to increased research activity in the areas of quantum computation and communications \cite{nielsen2010quantum, shor2000simple, steane1998quantum, walther2005experimental, barreiro2008beating, gisin2002quantum, gisin2007quantum}. In this regard, entangled photons (``biphotons'') are promising candidates due to their long coherence times and their capability to be projected into discretized \textit{d}-level entangled states in different degrees of freedom, such as time \cite{thew2004bell, Riedmatten2004Tailoring}, frequency \cite{olislager2010frequency, bernhard2013shaping, lukens2014generation,bessire2014versatile,xie2015harnessing}, orbital angular momentum \cite{Malik2016multi}, etc. More specifically, biphoton states in the form of a frequency comb (biphoton frequency comb, BFC) provide high-dimensionality in the frequency domain, and can be easily manipulated using electro-optic modulation and Fourier-transform pulse shaping \cite{kues2017chip, imany2017high} for quantum computation \cite{lukens2016frequency}. In addition, the frequency degree of freedom provides compatibility with standard optical fiber infrastructure and the ability to perform routing based on optical frequencies. However, showing that the photon-pairs are in a coherent superposition of frequency bins, is required for claims of frequency-bin entanglement.    

\begin{figure}[!b]
\centering
\includegraphics[width=\linewidth]{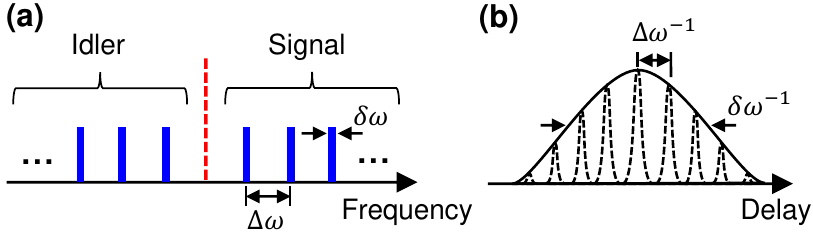}
\caption{Depiction of biphoton frequency comb (BFC).  (a) Spectrum of BFC with a free spectral range labeled as $\Delta\omega$. (b) Time correlation function, with fast substructure arising from coherent interference between the different biphoton frequency components. If the phase between different biphoton frequency components is random, there will be no time-average interference, and we would get only the longer envelope.}
\label{fig1}
\end{figure}

A straightforward approach to examine the coherence of a BFC is through temporal correlation measurements. If the two-photon spectrum is a coherent comb with a flat spectral phase, the temporal correlation would consist of a train of evenly spaced narrow peaks (see Fig. \ref{fig1}), which can be manipulated by adjusting the phase of different comb lines. In order to observe these features with a pair of single-photon detectors, the period of the correlation train would have to exceed the timing-jitter of the detectors; for example, a detection resolution of $\sim \ 100$  ps can only resolve the temporal structure of BFCs with a free spectral range (FSR) smaller than 10 GHz. And while nonlinear mixing techniques can be used for resolution improvement in coincidence measurements \cite{peer2005temporal, kuzucu2008time, lukens2014generation}, diminishing nonlinear efficiency makes this approach impractical for narrow-linewidth entangled photons. Nonetheless, electro-optic phase modulators can be employed to mix comb lines, which can then reveal spectral phase sensitivity even with slow single-photon detectors. In \cite{olislager2010frequency}, the authors used a pair of phase-modulators along with control of their modulation depths and relative phases to interfere biphotons, from which frequency entanglement was inferred; however, the input states to the ``two-photon interferometer'' had a continuous broadband spectrum and the notion of \textit{frequency-bins} was only implied from the application of narrow band spectral filters right before detection. Here we implement another phase-modulation scheme, as presented in \cite{imany2017two}, to demonstrate a proof-of-concept experiment, wherein our input states are BFCs obtained through spectral amplitude shaping of broadband biphotons; phase modulation in addition to spectral phase control enable us to observe high contrast interference fringes, a confirmation that the biphotons are indeed in a coherent superposition of frequency modes. Our new frequency domain scheme is in close analogy with Franson interferometry \cite{franson1989bell}, which has been widely applied in experiments on time-bin entangled photons \cite{thew2004bell, Riedmatten2004Tailoring}.

While a similar experimental setup has been explored in parallel for microresonator spontaneous four-wave mixing sources \cite{kues2017chip,imany2017high} which have narrow linewidth frequency bins ($\sim \ 100$ MHz), here we show that this phase modulation technique works well on the relatively wide frequency bins (12 GHz in our case) that are carved out of continuous, broadband spontaneous parametric downconversion spectra. Using a pulse shaper \cite{weiner2009ultrafast} along with the continuous broadband spectrum from downconversion gives us the ability to programmably carve out combs with a wide range of linewidths and spacings, unlike those generated through cavity-based parametric downconversion \cite{lu2003mode, scholz2007narrow}. Such biphoton frequency combs obtained through filtering of continuous spectra have been utilized in several recent experiments \cite{lukens2014generation,xie2015harnessing,bernhard2013shaping,bessire2014versatile}. Our results in conjunction with the spontaneous four-wave mixing works signify the universality of this approach for characterizing frequency-bin entanglement. Furthermore, we extend this technique to measure the dispersion in single-mode fibers using entangled photons. 

\section{II. Theoretical Background}

The state of a BFC can be written as 
\begin{equation}
\begin{gathered}
\label{eq:1}
\ket{\psi} = \sum\limits_{k=1}^N \alpha_k \ket{k,k}_{SI}, \\
\ket{k,k}_{SI} = \int d\Omega \, \Phi(\Omega-k\Delta\omega, \Omega+k\Delta\omega) \, \ket{\omega_0+\Omega, \omega_0-\Omega}_{SI},
\end{gathered}
\end{equation}
where $\ket{k,k}_{SI}$ indicates the $k^\textrm{th}$ mode (or comb line pair) of the signal and idler spectrum, $\alpha_k$ is a complex number representing the joint amplitude and phase of the $k^\textrm{th}$ comb line pair, $\Phi(\omega_\textrm{s},\omega_\textrm{i})$ is the lineshape of an energy-matched comb tooth pair, $\Delta\omega$ is the FSR, and $\omega_0=2\pi f_0$ is the center frequency of the biphoton spectrum. From here on, we will leave out the subscript, $SI$, from $\ket{k,k}_{SI}$.

Applying phase modulation of the form $\textrm{e}^{i\delta \sin \omega_\textrm{m}t}$  ($\omega_\textrm{m}$ is the modulation frequency and $\delta$ is the modulation depth) to a comb line projects it into sidebands offset from the original comb line by integer multiples of $\omega_\textrm{m}$ \cite{Harris2008nonlocal,Capmany2010quantum}---the positive-integer multiples correspond to upshifts in frequency while those of the negative-integers correspond to frequency downshifts. Thus, for a single photon, we can describe the effect of phase modulation on the $k^\textrm{th}$ frequency mode if in the signal and idler spectrum by 

\begin{align}
\label{eq:1b}
\begin{split}
\hat{m_\textrm{s}}\ket{k} = \sum \limits_{n=-\infty}^{\infty}C_n\Ket{k+\frac{n\omega_\textrm{m}}{\Delta\omega}}
\\
\hat{m_\textrm{i}}\ket{k} = \sum \limits_{m=-\infty}^{\infty}C_m\Ket{k-\frac{m\omega_\textrm{m}}{\Delta\omega}},
\end{split}
\end{align}
respectively, where $C_{n(m)}=J_{n(m)}(\delta)$ is the Bessel function which, when normalized, represents the probability amplitude of each frequency mode after phase modulation, and $J_{-n} = \textrm{e}^{in\pi}J_{n}$.
Consequently, the projection state of the $k^\textrm{th}$ biphoton mode after phase modulation of the signal and idler can be written as:
\begin{equation}
\label{eq:2}
\hat{m_\textrm{s}}\hat{m_\textrm{i}}\ket{k,k} = \sum \limits_{n,m=-\infty}^{\infty}C_nC_m \Ket{k+\frac{n\omega_\textrm{m}}{\Delta\omega}, k-\frac{m\omega_\textrm{m}}{\Delta\omega}}. 
\end{equation}
Therefore, we can project different comb line pairs into sidebands such that when they overlap, the emerging state would be in a superposition of indistinguishable frequency modes. 

As an example, let us consider two comb line pairs, $k$ and $k+1$, from the BFC and the first pair of sidebands $(n,m=\pm1)$ from phase modulation. After selecting only the sidebands that are in-between $k$ and $k+1$, in the signal and idler spectra, the biphoton state after projection can be written as:
\begin{equation}
\label{eq:3}
\begin{split}
\ket{\psi_\textrm{proj}} = \alpha_{k} & C_{1}C_{-1}\Ket{k+\frac{\omega_\textrm{m}}{\Delta\omega}, k+\frac{\omega_\textrm{m}}{\Delta\omega}} \\
&+ \alpha_{k+1}C_{-1}C_{1}\Ket{k+1-\frac{\omega_\textrm{m}}{\Delta\omega}, k+1-\frac{\omega_\textrm{m}}{\Delta\omega}}. 
\end{split}
\end{equation}
Now if we set $\omega_\textrm{m} = \Delta\omega/2$, the output state will become
\begin{equation}
\label{eq:4}
\ket{\psi_\textrm{out}} = C_{1}C_{-1}(\alpha_{k} + \alpha_{k+1})\Ket{k+\frac{1}{2}, k+\frac{1}{2}}.
\end{equation}
Hence, by selecting the resulting frequencies at $(k+\frac{1}{2})\Delta \omega$ from the center frequency, the two-photon coincidence rate, $\braket{\psi_\textrm{out}|\psi_\textrm{out}}$, truly originates from a superposition of contributions from the $k$ and $k+1$ frequency modes. Yet if the biphoton comb is coherent, we can observe two-photon interference in the coincidence rate by manipulating the phases of $\alpha_k$ and $\alpha_{k+1}$. 

\begin{figure*}[!tb]
\centering
\includegraphics[width=\textwidth]{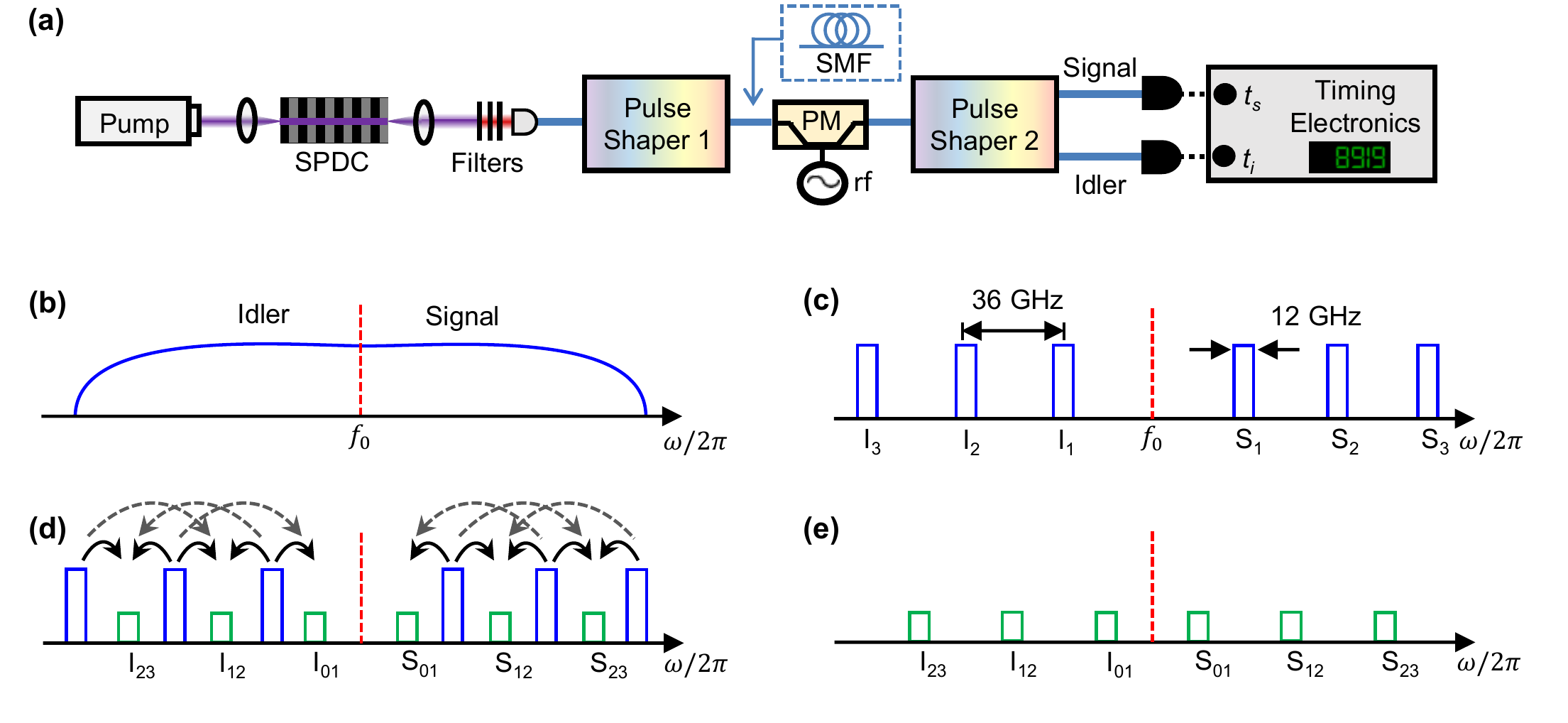}
\caption{Basic schematic for phase coherence measurements and illustration of biphoton spectral progression at different steps. (a) Experimental setup. (b) Broadband continuous biphoton spectrum. (c) Biphoton frequency comb after carving continuous spectrum with pulse shaper 1. The blocked frequencies were attenuated by 60 dB, making contamination from undesired frequencies negligible.  (d) Sidebands projected from phase modulation of comb lines. (e) Using pulse shaper 2, selected sidebands could be routed to a pair of single-photon detectors. SPDC: spontaneous parametric downconversion; PM: phase modulator; rf: radio frequency.}
\label{fig2}
\end{figure*}

Our experimental setup is presented in Fig. \ref{fig2}(a). We pump a 67-mm-long periodically poled lithium niobate waveguide with a continuous-wave laser at 771 nm in order to generate broadband biphotons centered around 1542 nm (194.55 THz). Figure \ref{fig2}(b) shows a conceptual picture of the broadband biphoton spectrum generated through spontaneous parametric downconversion; the signals are defined as photons in the higher frequency band while the lower-frequency photons are called idlers. After filtering out the pump photons, we couple the signal and idler photons into a commercial pulse shaper (pulse shaper 1, Finisar WaveShaper 1000s). Using pulse shaper 1, we carve the continuous broadband spectrum into a BFC with a linewidth of 12 GHz and an FSR of 36 GHz ($\Delta\omega/2\pi$) [Fig. \ref{fig2}(c)]. Pulse shaper 1 is also used to attenuate comb lines when necessary to ensure the amplitude equalization required for maximally entangled states \cite{bernhard2013shaping}, as well as applying spectral phase patterns to the signal and idler comb lines during measurements. Next, the BFC is sent into a phase modulator (EOSpace)---driven by an 18-GHz sinusoidal waveform (one-half the FSR of the BFC)---to create sidebands at integer multiples of 18 GHz (Fig. \ref{fig2}(d)). We then send the phase-modulated BFC into another pulse shaper (pulse shaper 2, Finisar WaveShaper 4000s), with which we pick out only overlapped sidebands that consist of projections from different signal and idler comb lines [Fig. \ref{fig2}(e)]. The selected sidebands from the signal and idler halves are sent to a pair of gated InGaAs single-photon detectors (Aurea SPD\_AT\_M2) and an event timer (PicoQuant HydraHarp 400) is used to record coincidences.

\section{III. Frequency-Bin Entanglement}

\begin{figure}
\includegraphics[width=\linewidth]{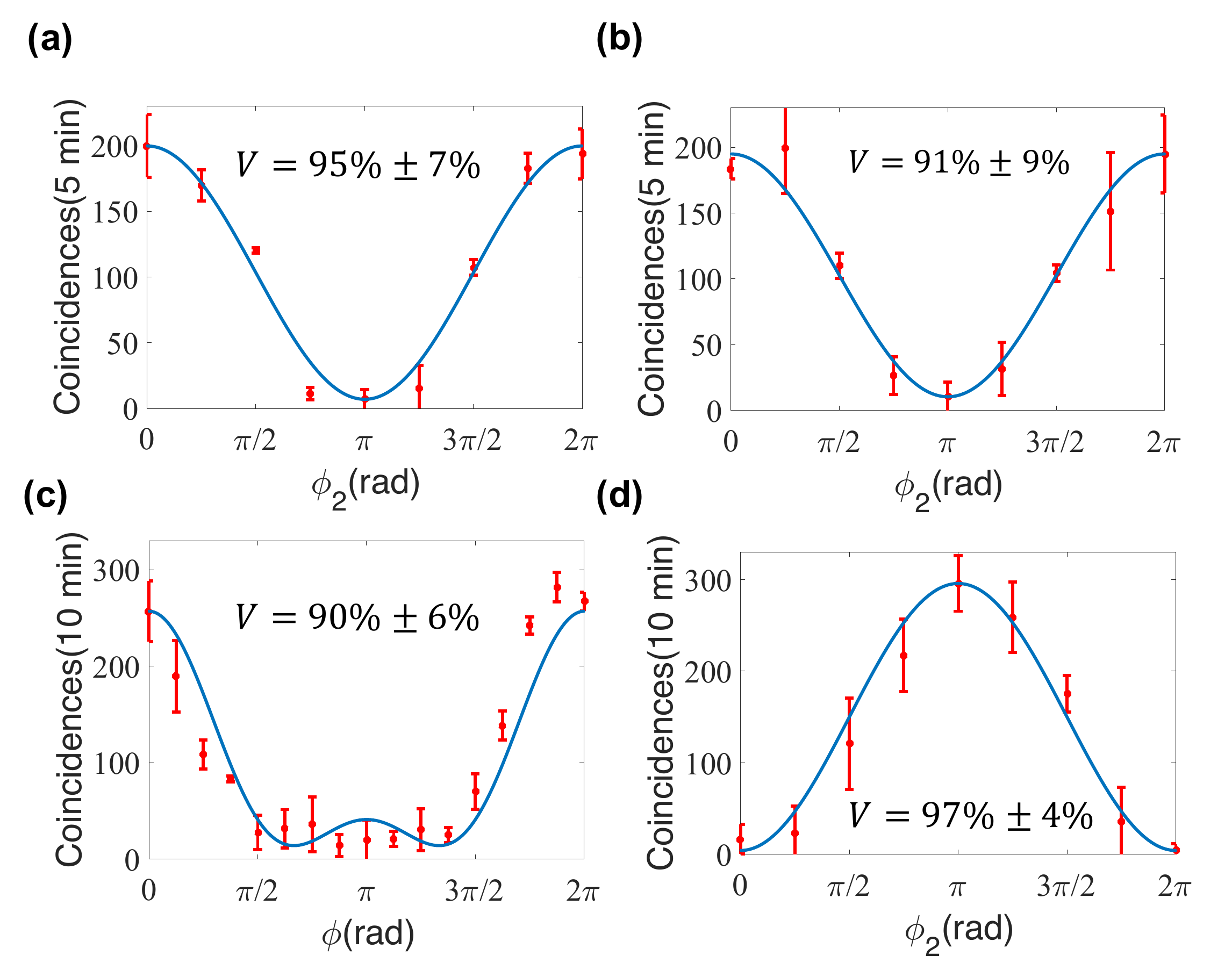}
\caption{Qubit and qutrit interference patterns. The two-photon interference as a result of applying (a) $\phi_2$ relative phase on $\textrm{S}_{2}\textrm{I}_{2}$ with respect to $\textrm{S}_{1}\textrm{I}_{1}$, (b) $\phi_2$ relative phase on $\textrm{S}_{3}\textrm{I}_{3}$ with respect to $\textrm{S}_{2}\textrm{I}_{2}$, (c) 0 phase on $\textrm{S}_{1}\textrm{I}_{1}$, $\phi$ phase on $\textrm{S}_{2}\textrm{I}_{2}$, and $2\phi$ phase on $\textrm{S}_{3}\textrm{I}_{3}$, (d) $\phi_2$ phase on $\textrm{S}_{2}\textrm{I}_{2}$ while setting the sideband amplitude such that $|C_3| = \frac{|C_1|}{2}$. The red error bars are the standard deviation of three measurements for each phase and the blue curves indicate the theoretical predictions taking into account the visibility calculated from the maximum and minimum data points. The coincidence-to-accidental ratio in our measurements was 3:1, but accidentals were subtracted in these plots.}
\label{fig3}
\end{figure}

For our first demonstration using this scheme, we create two comb line pairs, $\textrm{S}_1\textrm{I}_1$ and $\textrm{S}_2\textrm{I}_2$, while ensuring that the pairs contribute equal amplitudes ($|\alpha_1|^2 = |\alpha_2|^2$) by measuring coincidences between $\textrm{S}_1$ and $\textrm{I}_1$, and $\textrm{S}_2$ and $\textrm{I}_2$. We also apply a phase of $\phi_2/2$ to both $\textrm{S}_2$ and $\textrm{I}_2$, giving a total relative phase of $\phi_2$ on $\textrm{S}_2\textrm{I}_2$ with respect to $\textrm{S}_1\textrm{I}_1$. Then we drive the phase modulator with an rf power such that the frequency projection is mostly dominated by the first phase modulation sidebands, giving us $|C_{\pm1}|^2=0.32$ (the amplitude of each sideband is obtained by sending a continuous-wave laser through the phase modulator and measuring the output using an optical spectrum analyzer). After phase modulation, we pick out the overlapped sidebands---$\textrm{S}_{12}$ halfway between $\textrm{S}_1$ and $\textrm{S}_2$, and $\textrm{I}_{12}$ in the middle of $\textrm{I}_1$ and $\textrm{I}_2$. Sweeping $\phi_2$ from 0 to $2\pi$ and recording the coincidence rates, we obtain a sinusoidal interference pattern with a visibility of $95\% \pm 7\%$, shown in Fig. \ref{fig3}(a). The  pattern matches our expectation from theory, $\braket{\psi_\textrm{out}|\psi_\textrm{out}} \sim 1 + \cos \phi_2$, using Eq. (\ref{eq:4}) with $\alpha_{2}= \textrm{e}^{i\phi_2}\alpha_{1}$. Similarly, we repeated the experiment using comb line pairs $\textrm{S}_2\textrm{I}_2$ and $\textrm{S}_3\textrm{I}_3$, and picked out the overlapped sidebands in-between them $(\textrm{S}_{23}\textrm{I}_{23})$; the resulting interference pattern with a visibility of $91\% \pm 9\%$ is shown in Fig. \ref{fig3}(b). Thus we can confirm frequency-bin entanglement for the utilized $d=2$ states since the visibilities exceed $71\%$ \cite{thew2004bell}. Here we also note that the constructive and destructive interference points occur at $\phi_2 = 0$ and $\phi_2 = \pi$ respectively, suggesting that $\alpha_1\approx\alpha_2\approx\alpha_3$.

To explore $d=3$ frequency-bin entanglement, we utilize all three of the comb line pairs, $\textrm{S}_{1}\textrm{I}_{1},\  \textrm{S}_{2}\textrm{I}_{2}$, and $\textrm{S}_{3}\textrm{I}_{3}$ (setting $|\alpha_1|^2 = |\alpha_2|^2 = |\alpha_3|^2$). After phase modulation, we again pick out the sidebands $\textrm{S}_{12}$ and $\textrm{I}_{12}$, but in this case, $\textrm{S}_{12}$ consists of the sideband projections $n = 1,\ -1,\ -3$ from $\textrm{S}_{1},\ \textrm{S}_{2},\ \textrm{S}_{3}$, and $m = -1,\ 1,\ 3$ from $\textrm{I}_{1},\ \textrm{I}_{2},\ \textrm{I}_{3}$, respectively. We ensure that the magnitude of the first and third sidebands are equal by adjusting the rf power to give us $C_1= -C_{-1} = C_3 = -C_{3}$, and we measured $|C_1|^2$ to be 0.16. Now by applying a phase of 0 to comb line pair $\textrm{S}_{1}\textrm{I}_{1}$, $\phi$ to $\textrm{S}_{2}\textrm{I}_{2}$, and $2\phi$ to $\textrm{S}_{3}\textrm{I}_{3}$, the output state just before detection can be written as:
\begin{equation}
\label{eq:5}
\begin{split}
\ket{\psi_\textrm{out}} &= \alpha_1C_1C_{-1} + \alpha_2C_{-1}C_1\textrm{e}^{i\phi} + \alpha_3C_{-3}C_3\textrm{e}^{i2\phi} \Ket{\frac{3}{2}, \frac{3}{2}} \\
 &= -\alpha_1{C_1}^2(1+ \textrm{e}^{i\phi} + \textrm{e}^{i2\phi}) \Ket{\frac{3}{2}, \frac{3}{2}},
\end{split}
\end{equation}
if $\alpha_1 = \alpha_2 = \alpha_3$. The result obtained after sweeping $\phi$ from 0 to $2\pi$ is presented in Fig. \ref{fig3}(c).  Since we have contributions from three pairs of comb lines, the features of the interference pattern are now sharper compared to those observed in the Figs. \ref{fig3}(a) and \ref{fig3}(b); this sharpening is analogous to the sharpening of the pulses in a mode-locked laser as more frequency lines are added. We calculate a visibility of $90\%\pm6\%$, which is sufficient to prove entanglement between our entangled qutrits ($d=3$) since it is higher than the three-dimensional classical visibility threshold of $77.5\%$ \cite{thew2004bell}.

We can also manipulate the coincidence pattern resulting from the interference of three comb line pairs by looking at sidebands projected to other frequency locations as well as applying different phase configurations to the comb lines. Here we examine asymmetric sidebands, $\textrm{S}_{12}$ and $\textrm{I}_{23}$, containing contributions from the $n = 1,\ -1,\ -3$ sidebands of $\textrm{S}_{1},\ \textrm{S}_{2},\ \textrm{S}_{3}$, and $m = -3,\ -1,\ 1$ sidebands of $\textrm{I}_{1},\ \textrm{I}_{2},\ \textrm{I}_{3}$, respectively. Again, we set $\alpha_1 = \alpha_2 = \alpha_3$, but now we tune the rf power such that $|C_3| = \frac{|C_1|}{2}$ and then we apply a phase of $\phi_2$ to $\textrm{S}_{2}\textrm{I}_{2}$. The ensuing output state will be
\begin{equation}
\label{eq:6}
\begin{split}
\ket{\psi_\textrm{out}} &= \alpha_1C_1C_{-3} + \alpha_2C_{-1}C_{-1}\textrm{e}^{i\phi_2} + \alpha_3C_{-3}C_{1} \Ket{\frac{3}{2}, \frac{5}{2}} \\
&= -\alpha_1{C_1}^2\Big(\frac{1}{2} - \textrm{e}^{i\phi_2} + \frac{1}{2}\Big) \Ket{\frac{3}{2}, \frac{5}{2}}.
\end{split}
\end{equation}
Yet again we observe a sinusoidal interference pattern [Fig. \ref{fig3}(d)] when we sweep $\phi_2$ from 0 to $2\pi$, in agreement with theory---using Eq. (\ref{eq:6}), $\braket{\psi_\textrm{out}|\psi_\textrm{out}} \sim 1 - \cos \phi_2$. 

\begin{figure*}[t!]
\centering
\includegraphics[width=\linewidth]{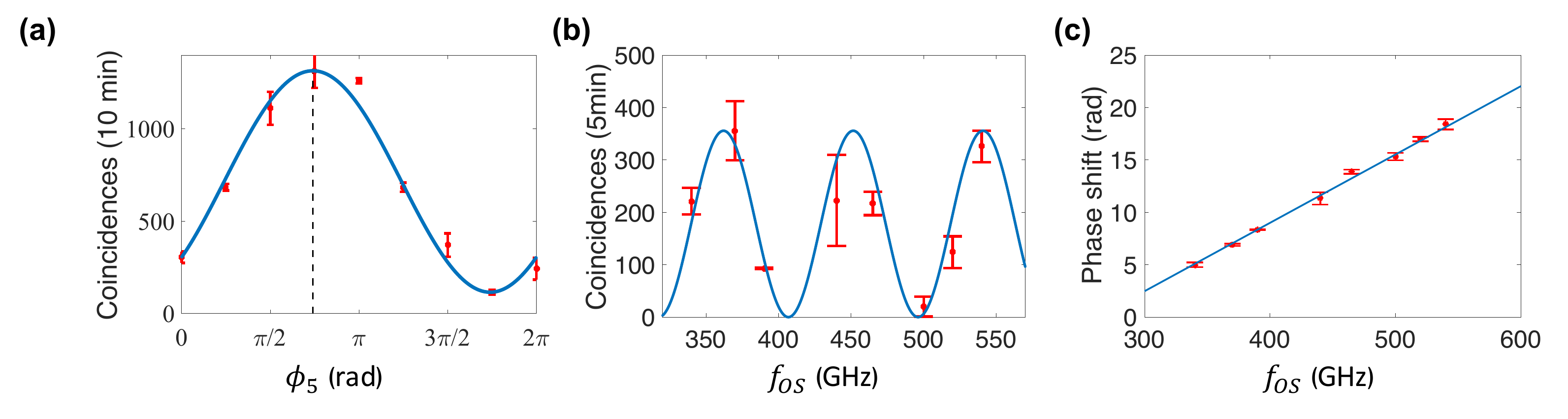}
\caption{(a) Shift in the interference pattern as a result of added dispersion; the dashed vertical line indicates the relative shift of $\phi_5=0.74\pi$. The blue curve indicates the theoretical prediction taking into account the visibility calculated from the maximum and minimum data points.(b) Coincidences as a function of $f_\textrm{os}$ when $\phi_{k}=\phi_{k+1}=0$. The blue curve is the theoretical prediction normalized to the maximum number of coincidence counts. (c) Phase shift of the interference pattern as a function of $f_\textrm{os}$.The blue line is the linear fit to the data points. The red error bars are the standard deviation of three measurements. The coincidence-to-accidental ratio was also 3:1 in these measurements and the accidentals were subtracted in the plots.}
\label{fig4}
\end{figure*}

\section{IV. Dispersion Measurement}

Moreover, the versatility of our experimental technique facilitates the measurement of dispersion using entangled photons. We  insert some SMF-28e fiber before pulse shaper 1 to induce dispersion on the biphotons [Fig. \ref{fig2}(a)]---the dispersion of this fiber around 1550 nm (extracted from the datasheet) is $D = 16.2 \, \textrm{ps}/(\textrm{nm} \ \textrm{km})$ and $\beta_2 = -D\lambda^2/2\pi c = -2.06 \times 10^{-2} \, \textrm{ps}^2/\textrm{m}$ \cite{weiner2009ultrafast}. Now we revisit the $d=2$ interference results shown in Figs. \ref{fig3}(a) and \ref{fig3}(b), and described by Eq. (\ref{eq:4}). Fiber dispersion will impart an additional relative phase on the ${(k+1)}^\textrm{th}$ bin with respect to the $k^\textrm{th}$, and this will lead to a phase shift in the interference pattern. The phase shift is given by
\begin{equation}
\begin{split}
\phi_\textrm{shift}&= -(2\pi)^2\beta_2l[(f_\textrm{os}+\Delta f)^2-f_\textrm{os}^2]\\
&=-(2\pi)^2\beta_2l\Delta f (2f_\textrm{os} + \Delta f) 
\end{split}
\label{eq:7}
\end{equation}
where  $l$ is the fiber length, $\Delta f=\Delta \omega/2\pi$ is the FSR in Hz, $f_\textrm{os}=k\Delta f$ is the frequency difference between the $k^\textrm{th}$ frequency bin and the center frequency, and we have assumed the dominant dispersion is the quadratic spectral phase term. [Unlike the classical term, a factor of 1/2 is dropped in Eq. (\ref{eq:7}) since the total phase shift is sum of relative phase shifts in the signal and idler comb lines.] As an initial experimental test, we use a fiber length of 200 m and select comb line pairs $\textrm{S}_5\textrm{I}_5$ and $\textrm{S}_6\textrm{I}_6$. Similar to previous measurements, after phase modulation, we pick out the sidebands $\textrm{S}_{56}$ between $\textrm{S}_{5}$ and $\textrm{S}_{6}$, and $\textrm{I}_{56}$ between $\textrm{I}_{5}$ and $\textrm{I}_{6}$, and then record the coincidence counts as we sweep $\phi_5$ from 0 to $2\pi$. The result, given in Fig. \ref{fig4}(a), shows a sinuosidal interference pattern albeit shifted by a phase of $0.74\pi$, in excellent agreement with theory [using Eq. (\ref{eq:7}) with $k=5$ and $\Delta f = 36$ GHz].  

For a complete frequency-dependent phase shift measurement, we replace the 200-m-long fiber with another fiber, 1.1 km long. However, rather than sweep $\phi_k$ for each $f_\textrm{os}$, we set it to zero and only register the coincidence counts as a function of $f_\textrm{os}$ [Fig. \ref{fig4}(b)]. We can then compute the phase shift for each $f_\textrm{os}$ by comparing its corresponding coincidence counts, $C(f_\textrm{os})$, to the expected maximum number of coincidences $C_\textrm{max}$. By measuring the same single photon count rates in the selected frequency bins, we ensure that $C_\textrm{max}$ is constant as a function of $f_\textrm{os}$. The phase shift will be given by $C(f_\textrm{os})=\frac{C_\textrm{max}}{2}[1+cos(\phi_\textrm{shift})]$, which we can unwrap to obtain the linear plot in Fig. \ref{fig4}(c). From Fig. \ref{fig4}(c), $\beta_2$ can be retrieved by calculating the slope of the curve [derivative of $\phi_\textrm{shift}$ with respect to $f_\textrm{os}$ in Eq. (\ref{eq:7})]. We obtain a value of $\beta_2=(-2.030\pm0.013)\times 10^{-2} \, \textrm{ps}^2/\textrm{m}$, not far off the $-2.06 \times 10^{-2} \, \textrm{ps}^2/\textrm{m}$ expected for SMF-28e fiber. 

\section{V. Conclusion} 
In conclusion, we have demonstrated a technique for verifying phase coherence in BFCs. The attributes of this approach, in which we mix adjacent frequency bins, are analogous to those of Franson interferometry, which mixes entangled photon time bins. Equivalently, our approach provides a straightforward path to prove frequency-bin entanglement; we presented interference patterns with visibilities higher than the classical threshold for entangled qubit and qutrit states. These results reinforce the potential of biphoton frequency combs as high-dimensional entangled states. Last, our dispersion measurements suggest the potential of low-light dispersion measurement with biphotons. 

\section{Acknowledgements}
This work was funded by the National Science Foundation under Grant No. ECCS-1407620. The authors thank C. Langrock and M. M. Fejer for fabrication of the PPLN waveguide. J. A. J. acknowledges support from Colciencias Colombia through the Francisco Jose de Caldas Conv. 529 scholarship and Fullbright Colombia.

P.I. and O.D.O. contributed equally to this work.

\bibliography{References.bib}
\end{document}